\documentclass[11pt,reqno,twoside]{article}
\usepackage{amsmath,amssymb,theorem,color}
\theorembodyfont{\itshape}
\newtheorem{thm}{Theorem}[section]
\newtheorem{lem}[thm]{Lemma}

{\theorembodyfont{\upshape}

}

\newcommand{\Proof}[1][]{\noindent{\itshape Proof#1. }}
\newcommand{\EndProof}{~$\Box$\bigskip}
\makeatletter
    
    \@addtoreset{equation}{section}
\makeatother

\def\mbR{{\mathbb R}}

\def\mbZ{{\mathbb Z}}

\def\mcB{\mathcal{B}}

\def\ol{\overline}

\def\sbs{\subset}

\def\ptl{\partial}

\def\a{\alpha}    \def\b{\beta}              
\def\D{\Delta}    \def\e{\varepsilon} \def\ph{\phi}       \def\Ph{\Phi}
             
\def\l{\lambda}        \def\m{\mu}        
\def\r{\rho}          \def\o{\omega}     \def\O{\Omega}
                   
\def\th{\theta}

\begin{document}
\title{Erratum: Dirichlet Forms and Dirichlet Operators\\
for Infinite Particle Systems: Essential Self-adjointness\footnote{This work appeared in {\it J. Math. Phys. } {\bf 39}(12), 6509-6536 (1998).}}
\author{Veni Choi\footnote{Institute for Mathematical Sciences, Yonsei University, 134 Shinchon-dong, Seodaemoon-gu, Seoul 120-749, Korea. E-mail: greennyel@gmail.com}, Yong Moon Park\footnote{Department of Mathematics, Yonsei University, 134 Shinchon-dong, Seodaemoon-gu, Seoul 120-749, Korea.  E-mail: ympark@yonsei.ac.kr}, and Hyun Jae Yoo\footnote{Department of Applied Mathematics,
Hankyong National University, 67 Seokjeong-dong, Anseong-si,
Gyeonggi-do 456-749, Korea. E-mail: yoohj@hknu.ac.kr} }
\date{}
%\subjclass{Primary: 60K35; Secondary: 82B21 }
  \maketitle

 In Lemma A.4, which was used in the proof of essential self-adjointness of the Dirichlet operator, it was erroneously stated that the space $D_0^2(\ol{\O})$ of certain local functions is invariant under the Markov semigroup $\{P_\r^t\}_{t\ge 0}$, that is given by solving the stochastic differential equations in (4.7). This lemma, and the other results of the paper hold in the present form for fintite range interactions. But in order to incorporate with infinite range interactions, we need to extend the core of the generator by relaxing the locality.

 It turns out that we need to modify the function spaces so that they reflect the decay rates of the considered interactions. We begin by introducing some convenient notations. First we will modify the space $C_1({\ol \O})$. For it, and for a later use, let us denote by $\mcB_0$ the class of nonnegative functions $a:\mbR_+\to \mbR$ such that

 (i) $0< a(0)$ and $a$ is increasing so that $a(\l)\to \infty$ as $\l\to \infty$;

 (ii) $\frac{\l}{a(\l)}$ is increasing.\\
 For example, $a(\l):=\a_l(\l)$ belongs to $\mcB_0$ for each $l\ge 1$, where $\a_l(\l)$ is a slowly increasing function that is defined below. We define
 \[
 C_1(\ol \O):=\{u\in C({\ol \O}):\, \text{there is an }a\in \mcB_0 \text{ and a constant }c_u \text{ s.t. }\|u\|_h\le \exp(c_u\,\frac{h}{a(h)}) \}.
 \]
We will consider some hierarchy for the subexponential order. Recall that $\a:\mbR_+\to \mbR$ is a monotonic increasing and concave function such that

(i) $\a(0)\ge 1$ and $\a(\l)\to \infty$ as $\l\to \infty$.

(ii) $\a'(\l)\le \frac{1}{1+\l}\a(\l)$ for $\l\ge 0$, and there exists a constant $c>0$ such that $\a''(\l)\ge -c\frac{1}{1+\l}$.

\noindent We define $\a_0(\l):=\a(\l)$ and inductively
\[
\a_l(\l):=\log(e+\a_{l-1}(\l)), \quad l=1,2,\cdots.
\]
It is not hard to check that for each $l\ge 0$, $\a_l(\l)$ satisfies the properties (i) and (ii) above. We modify the definition II.16 by \\[2ex]
{\bf Definition II.16':} {\it Let $D_{\text{se}}^q(\ol \O)$, $q=1,2,3$, denote the space of functions $u\in C_1(\ol \O)$ possessing derivatives of order less than or equal to $q$; each of these derivatives belong to $C_1(\ol \O)$. Moreover, there exist $\e\equiv \e_u>0$ and $l\equiv l_u\ge 2$ such that for any $h>0$, we have
\begin{eqnarray*}\label{eq:se_function}
\||D_ku(\o)|\exp\big(\e\log(1+|x_k|^2)\a_l(1+|x_k|^2)\big)\|_h\le c_u(\e,h)\text{ for }q=1,2,3   \\
\||D_jD_ku(\o)|\exp\big(\e\sum_{s=j,k}\log(1+|x_s|^2)\a_l(1+|x_s|^2)\big)\|_h\le c_u(\e,h)\text{ for }q=2,3\\
\||D_iD_jD_ku(\o)|\exp\big(\e\sum_{s=i,j,k}\log(1+|x_s|^2)\a_l(1+|x_s|^2)\big)\|_h\le
c_u(\e,h)\text{ for }q=3.
\end{eqnarray*}
Also for $q=1,2,3$, we define the spaces
\[
D_{\text{se}1}^q(\ol \O):=\{u\in D_{\text{se}}^q(\ol \O):\, \exists
a\in \mcB_0\text{ s.t. }  \text{ the above hold with
}c_u(\e,h)=\exp[c_u(\e)(1+\frac{h}{a(h)})]\}.
\]
}\\[2ex]
We will take $D_{\text{se}1}^2(\ol \O)$ as a defining domain of the Dirichlet operator. For it, we need the following
\begin{lem}
 For any $u\in D_{\text{se}1}^2(\ol{\O})$, $H_\m u\in L^2(\ol{\O},\m)$.
 \end{lem}
 \Proof Recall that for $\o=(x_k)_{k\in S}\in \ol{\O}$ (see (2.27))
 \begin{eqnarray*}
 H_\m u(\o)&=&-\frac12 \D u(\o)-\frac12\langle b(\o),\nabla u(\o)\rangle_{\o,0}\\
 &=&-\frac12\sum_{k\in S}\D_{x_k}u(\o)-\frac12\sum_{k\in S}\langle b_k(\o),D_{x_k}u(\o)\rangle_{\mbR^d}.
 \end{eqnarray*}
 Since $u\in D_{\text{se}1}^2(\ol{\O})$, we can find an $\e>0$, a constant $c_u(\e)$, $l\ge 2$, and $a\in \mcB_0$ such that the bound
 \[
 |D_{x_k}u(\o)|+|\D_{x_k}u(\o)|\le \exp[-\e \log(1+|x_k|^2)\a_l(1+|x_k|^2)]\exp[c_u(\e)(1+\frac{{\ol H}(\o)}{a({\ol H}(\o))})]
 \]
 holds. On the other hand, by the decreasing rates for the derivatives of the interaction given in the statement of Theorem II.17 we have the bound
 \begin{eqnarray*}
 |b_k(\o)|&=&|-\b \sum_{j\neq k}(\text{grad}\Ph)(x_k-x_j)|\\
 &\le &\b\sum_{j\neq k}\exp[-c_0\log(1+|x_k-x_j|^2)\a(1+|x_k-x_j|^2)].
 \end{eqnarray*}
Combining these we have the bound
\begin{eqnarray*}
|H_\m u(\o)|&\le & C\,\exp[c_u(\e)(1+\frac{{\ol H}(\o)}{a({\ol H}(\o))})]\Big[\sum_{r\in \mbZ^d}n(\o;r)\exp[-\e \log(1+|r|^2)\a_l(1+|r|^2)]\\
&& \times\Big(1+\sum_{s\in \mbZ^d:\,s\neq r}n(\o;s)\exp[-c_0\log(1+|r-s|^2)\a(1+|r-s|^2)]\Big)\Big]\\
&\equiv&C\, \exp[c_u(\e)(1+\frac{{\ol H}(\o)}{a({\ol H}(\o))})]\,A(\o),
\end{eqnarray*}
where $C$ is a constant. In order to see $H_\m u\in L^2(\ol{\O},\m)$, it is enough to check that both functions $\exp[c_u(\e)\frac{{\ol H}(\o)}{a({\ol H}(\o))}]$ and $A(\o)$ belong to $L^q(\ol{\O},\m)$ for any $q>1$. The fact that the function $A(\o)$ belongs to $L^q(\ol{\O},\m)$ promptly follows from Lemma III.1. Since $a(\l)\to \infty$ as $\l\to \infty$, in order to check that $\exp[c_u(\e)\frac{{\ol H}(\o)}{a({\ol H}(\o))}]$ belongs to $L^q(\ol{\O},\m)$, it is enough to show that
\[
\int \exp[\l \ol{H}(\o)]d\m(\o)<\infty
\]
for sufficiently small $\l>0$. But it is shown in (A16). This completes the proof of the lemma.
\EndProof\\[2ex]
Now in the statement of Theorem II.17, we replace $D_0^2(\ol \O)$ by $D_{\text{se}1}^2(\ol{\O})$. Since we use Proposition IV.1 for the proof of Theorem II. 17, we need also to replace $D_0^2(\ol \O)$'s by $D_{\text{se}1}^2(\ol{\O})$'s in the statements of Proposition IV.1.

The proof of Theorem II. 17 follows by using Proposition IV.2, Lemma IV.3, and the lemmas in the appendix. But, we also need slight modifications in the notations, though the proofs follow the same stream as before. Here we present them by naming with primes. We start by modifying Proposition IV.2. Recall that
\[
Z_1(t,z):=t+\int_0^t{\ol H}(\o(s,z,\r))^{1/2}ds.
\]
For each $l\ge 2$ we define
\[
F^{(l)}(t,z):=1+{\ol H}(\o(t,z,\r))^{1/2}+\frac{Z_1(t,z)^2}{\a_{l+1}(Z_1(t,z))^{1/2}}.
\]
For each $\ph\in D_{\text{se}1}^2(\ol \O)$, we let
\[
F_\ph(t,z):=\frac{{\ol H}(\o(t,z,\r))}{a({\ol H}(\o(t,z,\r)))}+F^{(l)}(t,z),
\]
where $a\in \mcB_0$ and $l\ge 2$ come from the defining property of $\ph$. \\[2ex]
{\bf Proposition IV.2':} {\it Suppose that the hypotheses of Theorem II.17 are satisfied and let $\ph\in D_{\text{se}1}^2(\ol{\O})$. Then there are constants $c>0$ and $K>0$ such that the bounds
\begin{eqnarray*}
&|\ol{D}_nP_\r^t\ph(z)|\le \exp[-c\log(1+|z_n|^2)\a_{l+1}(1+|z_n|^2)]\{E^W[\exp[KF_\ph(t,z)]]\}^{1/2},\\
&|\ol{D}_m\ol{D}_nP_\r^t\ph(z)|\le \exp[-c\sum_{s=m,n}\log(1+|z_s|^2)\a_{l+1}(1+|z_s|^2)]\{E^W[\exp[KF_\ph(t,z)]]\}^{1/2}
\end{eqnarray*}
hold uniformly in $\r\ge 1$, where $l\ge 2$ comes from the defining property of $\ph$. }\\[2ex]
{\bf Lemma IV.3':} {\it For any $\ph\in D_{\text{se}1}^2(\ol{\O})$, and for any $K>0$, $0\le t\le T$, we have
\[
\int_{\ol \O}E^W[\exp[KF_\ph(t,z)]]d\m(z)<C_\ph(T,K)<\infty,
\]
where $C_\ph(T,K)$ does not depend on $\r$.}\\[2ex]
The main idea is to use the modified $\th$-functions. For each $l\ge 1$, define
\begin{eqnarray*}
\th^{(l)}(x,\l)&:=&\exp\big[-\frac{\l}{\a_l(\l)^{1/2}}[1+\l^2+\log(1+|x^2|)\a_l(\sqrt{1+|x|^2})]^{1/2}\big]\\
&&\hskip 2 true cm \times \exp[-\log(1+|x|^2)\a_l(\sqrt{1+|x|^2})].
\end{eqnarray*}
Notice that the main difference of this new function from that of the original version is the last part of subexponentially decreasing term. Nonetheless, these class of functions have similar properties as the original one. Namely, we have \\[2ex]
{\bf Lemma A.1':} {\it There are positive constants $c_1$, $c_2$, $c_3$, and $c_4$, that may depend on $l$, such that the following hold:\\
(a) $-\frac{\ptl}{\ptl \l}\th^{(l)}(x,\l)\ge c_1(1+\log(1+|x|^2))^{1/2}\th^{(l)}(x,\l)$;\\
(b) $|\text{grad\,}\th^{(l)}(x,\l)|+|\D \th^{(l)}(x,\l)|\le c_2\th^{(l)}(x,\l)$;\\
(c) $\th^{(l)}(x,\l)\le c_3\th^{(l)}(y,\l)\exp[c_4\log(1+|x-y|^2)\a_l(1+|x-y|^2)]$.
 }\\[2ex]
 \Proof (a) All we have used in the proof of Lemma A.1 (a) is (A2) (ii), but the functions $\a_l$'s have the same property. We follow the methods used in the proof of Lemma A.1 (a). \\
 (b) We notice $\a_l(\sqrt{1+|x|^2})\le c''\log(e+|x|^2)$ for some constant $c''\equiv c''(l)$. By a direct calculation we obtain the result.\\
 (c) The proof is almost the same as that of Lemma A.1 (c). Without loss of generality we may assume $|y|\ge |x|$. To estimate the first half part of the ratio $\th^{(l)}(x,\l)/\th^{(l)}(y,\l)$, let $G(x,\l):=1+\l^2+\log(1+|x|^2)\a_{l}(\sqrt{1+|x|^2})$. By fundamental theorem of calculus,
 \begin{eqnarray*}
&& \Big|\frac{\l}{\a_{l}(\l)^{1/2}}G(y,\l)^{1/2}-\frac{\l}{\a_{l}(\l)^{1/2}}G(x,\l)^{1/2}\Big|\\
 &=&\frac{\l}{\a_{l}(\l)^{1/2}}\int_{|x|}^{|y|}\frac{d}{du}\big(1+\l^2+\log(1+u^2)\a_l(\sqrt{1+u^2})\big)^{1/2}du\\
 &\le &\int_{|x|}^{|y|}\frac{d}{du}\big(\log(1+u^2)\a_l(\sqrt{1+u^2})\big)du\\
 &=&\log(1+|y|^2)\a_l(\sqrt{1+|y|^2})-\log(1+|x|^2)\a_l(\sqrt{1+|x|^2}).
 \end{eqnarray*}
 The function $0\le u\mapsto \log(1+u^2)\a_l(\sqrt{1+u^2})$ is increasing and concave in the region $u\ge u_0$ for some constant $u_0>0$. Thus the last term is bounded by
 \[
 \log(1+(u_0+|y|-|x|)^2)\a_l(\sqrt{1+(u_0+|y|-|x|)^2}).
 \]
 Since the logarithmic function and $\a_l$ are concave we obtain the result. The second half part of $\th^{(l)}(x,\l)/\th^{(l)}(y,\l)$ is estimated by the same factor as seen from the above calculations.
 \EndProof \\[2ex]
 {\bf Lemma A.2':} {\it
 For any $l\ge 1$ and $\e>0$, there exist positive constants $c_5$ and $c_6(\e)$, that may depend on $l$, such that the bound
 \[
 \exp[-\e\log(1+|x|^2)\a_l(1+|x^2|)]\le \th^{(l+1)}(x,\l)\exp\big[c_5\frac{\l}{\a_{l+1}(\l)^{1/2}}(1+\l)+c_6(\e)\big]
 \]
 holds.
 }\\[2ex]
 \Proof
 We notice that
 \begin{eqnarray*}
 &&\th^{(l+1)}(x,\l)^{-1}\exp[-\e\log(1+|x|^2)\a_l(1+|x^2|)]\\
 &=&\exp\Big[ \frac{\l}{\a_{l+1}(\l)^{1/2}}\big[1+\l^2+\log(1+|x|^2)\a_{l+1}(\sqrt{1+|x|^2}) \big]^{1/2}\\
 &&\hskip 2 true cm -\log(1+|x|^2)\big[\e\a_l(1+|x|^2)-\a_{l+1}(1+|x|^2)\big]\Big]\\
 &\le&\exp\Big[ \frac{\l}{\a_{l+1}(\l)^{1/2}}\big[1+\l^2+\log(1+|x|^2)\a_{l+1}(\sqrt{1+|x|^2}) \big]^{1/2}\\
 &&\hskip 2 true cm-\frac12\e\log(1+|x|^2)\a_l(1+|x|^2)+c'\Big].
 \end{eqnarray*}
 We divide the $x$-$\l$ region into two subregions: $1+\l^2\le \log(1+|x|^2)(\a_l(1+|x|^2))^{1/2}$ and $1+\l^2> \log(1+|x|^2)(\a_l(1+|x|^2))^{1/2}$. First, in the region $1+\l^2\le \log(1+|x|^2)(\a_l(1+|x|^2))^{1/2}$, since $\a_{l+1}(\sqrt{1+|x|^2})\le c+(\a_l(1+|x|^2))^{1/2}$, the last expression is bounded by
 \[
 \exp\Big[c''+\log(1+|x|^2)(\a_l(1+|x|^2))^{1/2}-\frac12\e\log(1+|x|^2)\a_l(1+|x|^2)\Big]\le \exp(c_6(\e)),
 \]
 because $\a_l(1+\l^2)\to \infty$ as $\l\to \infty$. In the region $1+\l^2> \log(1+|x|^2)(\a_l(1+|x|^2))^{1/2}$, the quantity is bounded by
 \[
 \exp\big[c_5\frac{\l}{\a_{l+1}(\l)^{1/2}}(1+\l)\big].
 \]
 \EndProof\\[2ex]
 \Proof[ of Proposition IV.2'] The proof follows the former proof of Proposition IV.2, but we use the new $\th$-function. By a chain rule, we have
 \[
 \ol{D}_n^{(r)}\ph(\o)=\sum_{k\in S}\langle D_k\ph(\o),u_k\rangle_{\mbR^d},
 \]
 where $\o=\o(t,z,\r)\equiv(x_k(t,z,\r))_{k\in S}$ is the solution of (4.7) and $u_k\equiv \ol{D}_n^{(r)}x_k(t,z,\r)$. Since $\ph\in D_{\text{se}1}^2(\ol \O)$, there exist $l\ge 2$, $\e>0$, and an increasing function $a\in \mcB_0$ such that
 \[
 |D_k\ph(\o)|\le \exp[-\e\log(1+|x_k|^2)\a_l(1+|x_k|^2)]\exp\big[c_u(\e)(1+\frac{\ol{H}(\o)}{a(\ol{H}(\o))})\big].
 \]
 Therefore,
 \begin{eqnarray*}
|\ol{D}_n^{(r)}\ph(\o)|&\le &\exp\big[c_u(\e)(1+\frac{\ol{H}(\o)}{a(\ol{H}(\o))})\big]\sum_{k\in S}\exp[-\e\log(1+|x_k|^2)\a_l(1+|x_k|^2)]|u_k|\\
&\le &\exp\big[c_u(\e)(1+\frac{\ol{H}(\o)}{a(\ol{H}(\o))})\big]\Big(\sum_{k\in S}\exp[-\e\log(1+|x_k|^2)\a_l(1+|x_k|^2)]\Big)^{1/2}\\
&&\hskip 2 true cm \times \Big(\sum_{k\in S}\exp[-\e\log(1+|x_k|^2)\a_l(1+|x_k|^2)]|u_k|^2\Big)^{1/2}.
\end{eqnarray*}
As like in (A7), we can show $\sum_{k\in S}\exp[-\e\log(1+|x_k|^2)\a_l(1+|x_k|^2)]\le c_{13}\ol{H}(\o)^{1/2}$. We use Lemma A.2'. Then
\begin{eqnarray*}
|\ol{D}_n^{(r)}\ph(\o)|&\le &\exp\Big[c_u(\e)(1+\frac{\ol{H}(\o)}{a(\ol{H}(\o))})+c_{13}\ol{H}(\o)^{1/2}+K_2 \frac{Z_1}{\a_{l+1}(KZ_1)^{1/2}}(1+Z_1)+K_2\Big]\\
&&\hskip 2 true cm \times R_1^{(l)}(t,KZ_1)^{1/2},
\end{eqnarray*}
where $R_1^{(l)}(t,\l):=\sum_{k\in S}\th^{(l+1)}(x_k(t,z,\r),\l)|u_k(t,z,\r)|^2$. Following the proof of (A10) we can show that for large values $K>0$,
\[
E^W[R_1^{(l)}(t,KZ_1(t,z))]\le R_1^{(l)}(0,0)=\exp\big[-\log(1+|z_n|^2)\a_{l+1}(1+|z_n|^2)\big].
\]
The proof of the first part of the proposition is completed. For the proof of second part we use the method employed in the proof of Proposition 6 of Ref. 7 together with necessary bounds in Lemma A.3. In Lemma A.3, which hold in that form, the function $h$ is $\a_1$ in the present notation. $\square$\\[2ex]
{\bf Proof of Lemma IV.3':} By Schwarz inequality it is enough to
show the inequalities separately:
\begin{eqnarray*}
&\int_{\ol \O}E^W\big[\exp[K(\frac{\ol{H}(\o(t,z,\r))}{a(\ol{H}(\o(t,z,\r)))}+\ol{H}(\o(t,z,\r))^{1/2})]\big]d\m(z)<C_a(T,K);\\
&\int_{\ol \O}E^W\big[\exp[K
\frac{Z_1(t,z)^2}{\a_l(KZ_1(t,z))^{1/2}}]\big]d\m(z)<C_l(T,K).
\end{eqnarray*}
Notice that for any $l\ge 2$, the function $\a_l$ has the similar
behavior as $h$ in (A14b), in particular, the function $0\le
\l\mapsto \l^2/\a_l(\l)^{1/2}=\l^2/\sqrt{\log(e+\a_{l-1}(\l))}$ is
convex. Therefore the second bound follows as in the proof of Lemma
IV.3. For the first inequality we use the invariance of $\m$ w.r.t.
$P_\r^t$ again. Then it reduces to show
\[
\int_{\ol \O}E^W\big[\exp[K(\frac{\ol{H}(z)}{a(\ol{H}(z))}+\ol{H}(z)^{1/2})]\big]d\m(z)<C_a(T,K).
\]
Since $\frac{\ol{H}(z)}{a(\ol{H}(z))}+\ol{H}(z)^{1/2}=\big(\frac{1}{a(\ol{H}(z))}+\frac1{\ol{H}(z)^{1/2}}\big)\ol{H}(z)$, and since the function $\l\mapsto \frac{1}{a(\l)}+\frac1{\l^{1/2}}$ goes to zero as $\l\to \infty$, it is again enough to show that for sufficiently small $\l>0$,
\[
\int_{\ol \O}\exp(\l\ol{H}(\o))d\m(\o)<\infty,
\]
which was shown in (A16). \EndProof\\[2ex]
{\bf Lemma A.4':} {\it For any $\r\ge 1$, $t\ge 0$,
$P_\r^t(D_{\text{se}1}^2(\ol \O))\sbs D_{\text{se}1}^2(\ol
\O)$.}\\[2ex]
\Proof Recall $u_\r(t):=P_\r^t\ph(z)=E^W[\ph(\o(t,z,\r))]$ for a
given $\ph\in D_{\text{se}1}^2(\ol \O)$. We first check that $C_1(\ol \O)$ is invariant under the semigroup $\{P_\r^t\}_{t\ge 0}$. Notice that by (4.7) the particles outside the ball $B_{2\r}(0)$ of radius $2\r$ centered at the origin are frozen, and the number of particles in $B_{2\r}(0)$ are conserved. By the superstability and the decay property of the interaction
$\Ph$, one can check that there exists a positive constant $c(\r)$
such that the bound
\[
\ol{H}(\o(t,z,\r))\le c(\r)\ol{H}(z), \quad 0\le t\le T,
\]
holds. From the above bound, it is easy to check that if $\ph$ belongs to $C_1(\ol \O)$ then $P_\r^t\ph\in C_1(\ol \O)$. Next suppose that $l\ge 2$ and
$a\in \mcB_0$ are respectively the number and increasing function
for $\ph$ satisfying the defining properties. By Proposition IV.2',
we have the bounds:
\[
|\ol{D}_nu_\r(t)|\le
\exp[-c\log(1+|z_n|^2)\a_{l+1}(1+|z_n|^2)]\{E^W[\exp[KF_\ph(t,z)]]\}^{1/2}
\]
and
\[
|\ol{D}_m\ol{D}_nu_\r(t)|\le
\exp[-c\sum_{s=m,n}\log(1+|z_s|^2)\a_{l+1}(1+|z_s|^2)]\{E^W[\exp[KF_\ph(t,z)]]\}^{1/2},
\]
where
\[
F_\ph(t,z)=\frac{{\ol H}(\o(t,z,\r))}{a({\ol H}(\o(t,z,\r)))}+1+\ol{H}(\o(t,z,\r))^{1/2}+\frac{Z_1(t,z)^2}{\a_{l+1}(Z_1(t,z))^{1/2}}.
\]
Notice that the functions $\l/a(\l)$, $\l^{1/2}$, and
$\l^2/\a_{l+1}(\l)^{1/2}$ are increasing. Then, by the bound $\ol{H}(\o(t,z,\r))\le c(\r)\ol{H}(z)$ given above, and also by noticing $Z_1(t,z)\le C(T,\r)\sqrt{1+{\ol H}(z)}$, we have
\[
\{E^W[\exp[KF_\ph(t,z)]]\}^{1/2}\le
\exp\Big[K'(T,\r)\big[\frac{\ol{H}(z)}{a(\ol{H}(z))}+1+\ol{H}(z)^{1/2}+\frac{\ol{H}(z)}{\a_{l+1}(\sqrt{1+\ol{H}(z)})^{1/2}}\big]\Big].
\]
Since $\l^{1/2}\le 1+\frac{\l}{\max\{1,\l^{1/2}\}}$, we may bound the r.h.s. of the above display by 
\[
\exp\Big[2K'(T,\r)\big[1+\frac{\ol{H}(z)}{a'(\ol{H}(z))} \big]\Big],
\]
where $a'(\cdot)\in \mcB_0$ is defined by $\frac{1}{a'(\l)}=\frac{1}{a(\l)}+\frac1{\max\{1,\l^{1/2}\}}+\frac1{\a_{l+1}(\sqrt{1+\l})^{1/2}}$. This completes the proof.   
\EndProof\\[2ex]
\Proof[ of Theorem II.17] By Lemma A.4', we have
$P_\r^t(D_{\text{se}1}^2(\ol \O))\sbs D_{\text{se}1}^2(\ol \O)$, and
in particular, $u_\r(t)\in D(H_\m)$. It remains to show the
condition (c) in Proposition IV.1. It follows from the same method
that was done in the proof of the original version of Theorem II.17.
\EndProof\\[3ex]
{\it Acknowledgement:} We are grateful to Eugene Lytvynov for
pointing out this error.

\end{document}